\documentstyle[aps,twocolumn]{revtex}

\catcode`\@=11

\begin{document}
\title{Dewetting, partial wetting and spreading of a two-dimensional monolayer
on solid surface}

\vspace{2cm}

\author{G.Oshanin$^{1,2}$,
  J.De Coninck$^1$, A.M.Cazabat$^3$ and M.Moreau$^2$}

\vspace{3cm}

\address{$^1$ Centre de Recherche en Mod\'elisation Mol\'eculaire, 
Universit\'e de Mons-Hainaut, 20 Place
du Parc, 7000 Mons, Belgium\\
$^2$ Laboratoire de Physique Th\'eorique des Liquides,
Universit\'e Paris VI, 4 Place Jussieu, 75252 Paris Cedex 05, France\\
$^3$ Laboratoire de Physique de la Mati\`ere Condens\'ee,
Coll\`ege de France, 11
Place M.Berthelot, 75231 Paris Cedex 05, France}

\vspace{3cm}

\address{\rm (Received: December 22, 1998)}
\address{\mbox{ }}
\address{\parbox{14cm}{\rm \mbox{ }\mbox{ }
We study the behavior  
of a semi-infinite monolayer, 
which is placed initially on a half of 
an infinite in both directions, ideal crystalline surface,  
and then evolves in time  due to random motion of the monolayer
particles. Particles  dynamics is  modeled as
the Kawasaki particle-vacancy 
exchange process in the presence of
long-range attractive particle-particle 
interactions.
In terms of an analytically solvable 
mean-field-type approximation
we calculate the 
mean displacement $X(t)$ of the 
monolayer edge
and discuss the 
conditions under which a monolayer spreads ($X(t) > 0$), 
partially wets ($X(t) = 0$)
or dewets from the solid surface ($X(t) < 0$).
}}
\address{\mbox{ }}
\address{\parbox{14cm}{\rm PACS No: 68.15 +e; 05.60 +w; 64.60.Ht; 64.90. +b}}
\maketitle

\makeatletter
\global\@specialpagefalse
\makeatother


Dynamics and static properties
 of thin liquid films on
 solid surfaces have been studied for
many years resulting in
 a seemingly good  understanding of
 the problem  \cite{pgg,adam}.
However, 
with the advent of new experimental techniques,
capable of studying properties of molecularly thin (MT) films,
 it has become clear that the developed theoretical concepts
 apply only to sufficiently thick films; for MT
films significant departures from the standard behavior
have been observed \cite{gran,caza}.
In particular, several remarkable features
 have been revealed by
ellipsometric
studies of the MT precursor films, i.e. 
films emitted by (sessile) liquid drops placed on
solid substrates \cite{caza}: \\
First, such films have been detected even in the 
case of non-wetting  
drops. This implies that physical conditions at which
such a MT film appears may be different of the ones corresponding
to
 the  
wetting/dewetting  transition at $\em macroscopic$ scales. 
Next, precursors do not spread at a constant
rate; the mean displacement of the film's 
edge grows with time $t$ only in 
proportion to $\sqrt{\rm t}$.
Lastly, "fine structure" of the MT precursors
may be very different; in some cases
the film's density 
shows a  pronounced variation with the distance from the
macroscopic drop, which reveals 
the surface-gas-like, rather than the
liquid-like behavior. In other systems, 
the films are  dense and compact.
Even more striking,  
 on 
the intermediate-energy
substrates surprising "terraced" patterns appear, 
formed  by 
several superimposed MT precursors each
spreading at the $\sqrt{\rm t}
$-rate 
on top of lower layers.

Meanwhile, several attempts have been made to explain
why do the MT precursor films spread
at the $\sqrt{\rm t}
$-rate.  
Ref.5 proposed a "stratified droplet" model, 
in which 
a sessile drop is regarded
as a succession of horizontal layers, each layer being
a two-dimensional, incompressible Navier-Stokes liquid. 
This model suggests that the $\sqrt{\rm t}
$-law
 results from the competition between
the liquid-solid attractions, which represent
the driving force of spreading,
and viscous-type frictional forces, which 
control particles  dynamics on the
solid surface. We  note parenthetically that
similar ideas have been used
to describe  
dynamics of the reverse processes - dewetting of a monolayer \cite{pgg1} and 
squeezing of a MT lubricating film 
out of a  gap between two
solids \cite{tos}, for which it has been also predicted  that
the
radii of the dewetted areas grow at the
$\sqrt{\rm t}
$-rate. 
Next, Ref.8 described  droplet spreading
in terms of 
the Langevin dynamics of a
non-volatile
fluid edge, modeled by horizontal
 solid-on-solid-model strings.
Such an approach has reproduced
the 
"terraced" 
profiles; 
the $\sqrt{\rm t}
$-law was found,
 however, only as
a transient regime. 
Lastly, in a microscopic approach of Refs.9  the MT
film was considered
 as a lattice gas of 
interacting particles connected
to a reservoir (droplet). 
Here,
the $\sqrt{\rm t}
$-law was obtained
 for both  precursors of the
sessile drops and creeping films in the 
capillary rise geometries;
 it was claimed
that such a  behavior
 is controlled by migration of voids
from the advancing
edge of the  film to the reservoir.

Despite reasonably good explanation 
of the dynamical behavior, provided by Refs.5,8 and 9,
several fundamental
questions still remain
largely unanswered. 
In particular,  
the dependence of  the
 prefactor in the $\sqrt{\rm t}
$-law  on 
the temperature, $\rm k_{B}  T$, 
and on the parameters of the
interaction potentials has not been elucidated so far. 
As a matter of fact, 
the model of Ref.5 discards 
 the effects of the drop's surface 
tension and/or 
of the monolayer edge tension $\gamma_{e}$ 
on spreading kinetics. 
In consequence, 
 Ref.5 predicts that "terraced" spreading 
appears as soon as any kind of attractive liquid-solid
interactions (LSI) is present, which 
contradicts apparently to the experience \cite{caza}.
Contrary to Ref.5, the models of Refs.8 and 9 take $\gamma_{e}$ into account
and 
show that the film 
may actually appear
only if the strength of the LSI
exceeds certain threshold value. 
However, a common subtle point 
of both Refs.8 and 9 is that
the prefactor in the $\sqrt{\rm t}
$-law is
expressed in terms of several parameters, which are
assumed to be independent of the dynamics;  in Refs.9,
for instance, these are
the 
particle density in the reservoir
and $\gamma_{e}$.  On the other hand, 
 $\gamma_{e}$
 originates from 
attractive
liquid-liquid interactions (LLI)
 and thus depends on the density profile
in the film. The latter 
is itself dependent on the 
spreading rate and
hence, on $\gamma_{e}$. 
Therefore, the calculation of $\gamma_{e}$ 
and, consequently,
of the prefactor in the $\sqrt{\rm t}
$-law  
requires solution
of essentially non-linear dynamical problem in which
attractive LLI 
are taken into account
 explicitly.

In this Rapid Communication we study analytically 
the 
behavior of a 
liquid monolayer,
which  occupies initially
a bounded, macroscopically large area of the solid surface,
and  then  evolves in time 
due to random motion
of the monolayer particles. Particles'  dynamics is modeled as
the Kawasaki-type particle-vacancy exchange process in the presence
of short-range repulsive (hard-core) and weak long-range 
attractive particle-particle interactions. 
Here we consider 
a simple case when
the initially occupied region 
 is the half-plane 
$- \infty < X \leq 0$, Fig.1, and  calculate 
the mean displacement $X(t)$
of the monolayer edge. 
We note that our results apply, as well, 
to the intermediate time
behavior in several other two-phase geometries. Particularly, 
the initially dewetted region can be a hole of radius $R$, 
nucleated in a homogeneous monolayer,
or the monolayer can occupy a 
circular region of radius $R$, which  situation appears
 at the late stages of sessile drops spreading \cite{caza}.
For such geometries,
 our results  describe the 
kinetics on time scales such that $X(t) \ll R$, in which regime
the precise form of the phase-separating 
boundary is not important (see, e.g. Refs.9).

To determine the time evolution of $X(t)$, we develop 
a mean-field-like, 
self-consistent approach, in which 
the  non-linear coupling
between the density distribution in the spreading 
film and the edge tension $\gamma_{e}$
is taken into account explicitly. Within this approach we 
 recover the result of Refs.9, i.e. the law $X(t) = 
\rm A \sqrt{\rm D_{0} t}$,
in which $\rm D_{0}$ 
is the bare diffusion coefficient
describing dynamics of an isolated particle on the solid surface.
Here, however, 
we define the prefactor $\rm A$ explicitly 
as a function
of $\rm k_{B}  T$
and of the interaction parameters. We show that $\rm A$ 
can be positive or negative, which means
that the monolayer can spread, 
partially wet or dewet from the solid surface, 
and determine the 
temperature $\rm T_{\rm w/dw}$
of the wetting/dewetting transition 
in the monolayer regime. 
Moreover, we find that spreading of the monolayer 
can proceed quite differently at different $\rm T$,
which agrees with the
experimentally observed behavior \cite{caza}.
When $\rm T \geq  \rm T_{\rm b}$, where $\rm T_{\rm b}$
is also found explicitly, we have 
that $\gamma_{e} \equiv 0$, 
$\rm A 
\sim \sqrt{ln(t)}$ as $\rm t 
\to \infty$,
and the density in the 
monolayer 
varies strongly 
with the distance from the edge. 
We remark that 
this finding contradicts to  
 Refs.9, which suggest that such 
a "surface-gas"-like 
spreading may take place 
only in the absence of attractive LLI.
For lower $\rm T$, 
such that $\rm T_{\rm w/dw} \leq T < T_{\rm b}$,
we find that 
the density variation is less pronounced and
both $\rm A$ 
and $\gamma_{e}$ 
are positive and constant, which signifies
that in this $\rm T
$-range the monolayer spreads as a "liquid". Lastly, 
for $\rm T < \rm T_{\rm w/dw}$
the monolayer dewets from the surface.

The model to be studied here is defined as follows:\\ 
(a) The  particles
experience two types of 
interactions - the LLI
and  the 
LSI. The LSI create effectively a lattice
 of potential wells
(with the coordination number $z$ and the interwell distance $\sigma$), 
such that the 
particles reside  in the local minima of these wells.
We assume that the LSI correspond to 
the limit of the so-called intermediate
localized adsorption, which  is appropriate for many adsorbates and
persists over a wide $\rm T
$-range \cite{adam}.
In this limit the particles are neither completely fixed in the wells,
nor completely mobile: The LSI wells are  deep with respect to desorption
(desorption barrier $\rm{U}_{d} \gg \rm k_{B} T$) 
so that only a monolayer can exist, but 
have much lower barrier $\rm{U}_{l}$ against 
the  movement across the surface.
Further on, we suppose that the LLI are  two-body,
central and additive;  the
LLI potential $\rm{U(r)}$ is a hard-core
at distance $\rm{r}
 = \sigma$, which means that each well can be occupied by one particle at most,
and is attractive for $r > \sigma$,
 $\rm{U(r)} =  - {\rm U}
_{0}(T) (\sigma/r)^{n}$, $n
> 2$. The parameter $\rm{U}
_{0}(T) \ll \rm{U}_{d}$, 
which is  the case for many realistic situations \cite{adam} and which
implies that the LLI 
incur only  small local perturbations to the array of the LSI wells.\\
(b) Occupation 
of the well
with radius-vector $\bf r$ at time $\rm t$ 
is described by the variable
$\eta(\bf r; \rm t)$, which can assume
 two values, $0$ and $1$;
its realization average
value, i.e. the local density, 
is denoted as $\rho(\bf r; \rm t) = 
\overline{\eta(\bf r; \rm t)}$.
The initial configuration of the monolayer is 
depicted in Fig.1, i.e. the monolayer
 particles are placed at random positions
and at a fixed coverage $\rho < 1$
(number of occupied wells as a fraction of the  total 
number of wells per unit area)
in the wells of the half-plane $- \infty < X \leq 0$.\\
(c) The particles motion is activated by chaotic vibrations of solid atoms
and proceeds by rare events of hops between the local minima of adjacent wells.
In absence of the LLI, one may  estimate  the diffusion coefficient
 for such a motion to be
$\rm{D}
_{0} \approx \omega \sigma^{\rm 2}/z$, where $\omega$
is
the  frequency of 
 hops.
   Now, the LLI couple the dynamics of any given particle
to the motions of all other particles: first, hard-core repulsion 
prevents multiple occupancy of any potential well;
second, on escaping 
 from the well with radius-vector $\bf{r}$,  
any given particle follows preferentially 
the local gradient of the  LLI potential landscape
$\rm{U}
(\bf{r};
{\rm t})$:
\begin{equation} 
{\rm U}
(\bf{r};
{\rm t}) \; = \; - \; 
{\rm U}
_{\rm{0}}
(\rm{T}) \; \sigma^{n} \; \sum_{\bf r''} 
\frac{\eta(\bf{r''};
{\rm t})}
{|\bf{r} - \bf{r''}|
^{n}},  
\end{equation}
where the summation extends 
over the entire lattice, excluding ${\bf r''} = {\bf r}$.
To take the LLI into account, we stipulate 
 that for any given particle
leaving 
the well with radius-vector $\bf{r}$ at time moment $\rm t$, 
the choice of the jump 
direction is random
and  governed by the position- and
 time-dependent probabilities \cite{leba}:
\begin{equation} 
p(\bf{r}|\bf{r'_{z}}) 
\; = \; {\rm Z}^{\rm{-1}} \;   
\exp( \frac{\beta}{\rm{2}} \;
 [\rm{U}
(\bf{r};
\rm{t})  
-  \rm{U}
({\bf{r'_{z}}};
{\rm t})]), \; 
\end{equation}
where $\bf{r'_{z}}$ is the radius-vector of 
one of $z$ wells, adjacent to the well at  $\bf{r}$, 
$\beta = 1/\rm k_{B} T$,
 and 
${\rm Z}$ denotes the normalization factor. 
 When the
 jump direction 
is chosen, the particle attempts to 
hop into the target well.
The hop is fulfilled if this well is unoccupied; 
otherwise, the particle attempting to hop
is repelled back to the well at $\bf{r}$.

We proceed further on by assuming local equilibrium (see, e.g.  Refs.10), 
which implies that 
occupations of different
wells factorize and thus allows 
for the description in 
terms of local densities, 
$\rho(\bf{r};\rm{t})$. For these,  we find 
\begin{eqnarray}
\frac{1}{\omega} 
\; \frac{\partial }{\partial t} 
\; \rho(\bf{r};\rm{t}) \; = \;
- \; \rho(\bf{r};\rm{t}) \; & \sum_{\bf r'_{z}} &
 \overline{p(\bf{r}|\bf{r'_{z}})} 
\; (1 \; - \; \rho(\bf{r'_{z}};\rm{t})) \; + 
\medskip
\nonumber \\
+ \; (1 \; - \; \rho(\bf{r};\rm{t})) \; 
& \sum_{\bf r'_{z}} & \overline{p(\bf{r'_{z}}|\bf{r})} \;
\rho(\bf{r'_{z}};\rm{t}),
\end{eqnarray}
in which $\overline{p(\bf{r}|\bf{r'_{z}})}$ are determined by
 Eqs.(2) and (1) with 
$\eta(\bf{r};\rm{t})$ 
replaced by $\rho(\bf{r};\rm{t})$.
Eq.(3) has to be solved subject to the initial
condition that $\rho(\bf{r};\rm{t}=0) = 0$ for $X > 0$
and $\rho(\bf{r};\rm{t}=0) = \rho$ for $X \leq 0$.

Let us analyse now, on the basis of  Eq.(3), 
the time evolution of the mean displacement $X(t)$
of  the 
monolayer edge.  To do this, we follow  
 Refs.9 supposing that for the  
long-ranged, but rapidly vanishing LLI,
a hop of any particle which is not 
 directly at the edge, does not
change  the energy in Eq.(1). 
This means, in virtue of Eq.(2), that
for such particles all hopping directions are equally probable
and their migration on the surface is
 constrained by the hard-core interactions only. 
The particles being at the
edge, however, are effectively attracted by
the "bulk" monolayer, which
results in asymmetric hopping probabilities
 (see Refs.9 for more details).
We note that such an approximation is actually 
a translation onto the 
molecular level of standard descriptions of the
 liquid front dynamics  as
a  competition between  surface tension 
and  internal pressure \cite{pgg,adam}.

Further on, since
we are interested in the behavior 
of the mean displacement of the edge, 
it is justified to neglect the fluctuations around $X(t)$ 
along the $Y$-axis. Consequently, we suppose  that
the edge position 
 does 
not depend on $Y$, Fig.1,  which 
makes the system effectively one-dimensional and the 
original two-dimensional geometry 
enters only
through the 2D diffusion coefficient $\rm D_{0}$
and  2D edge tension, (see Eq.(9)).

Now, the one-dimensional model  
we have to study consists of a 1D
hard-core lattice gas, which 
is put
 initially 
 into a "shock"
 configuration and then
evolves in time  by particles attempting
to hop to the nearest
unoccupied sites. The "shock" 
configuration
means that 
all particles are placed at random
with a fixed  
mean density $\rho$ at the sites 
$- \infty < X \leq 0$ 
of an infinite in
both directions 1D lattice.
All particles,
except the rightmost one (RMP) which defines position of the edge,
have equal probabilities ($= 1/z$) for hopping to 
the left or to the right.
The RMP, whose position is $X(t)$, is
attracted by the gas particles and thus
 has asymmetric hopping probabilities,
which obey Eq.(2) with 
$\eta(\bf{r};\rm{t})$ 
replaced by $\rho(\bf{r};\rm{t})$.

To determine $X(t)$ 
we now proceed as follows. Recollecting first the results
 of Refs.9, 
we anticipate
 that at sufficiently
large times the particle density 
past the RMP tends to
some constant value.
This implies, in virtue of Eqs.(1) and (2),
that  $p(X(t)|X(t) \pm \sigma)$ approach 
limiting values $p_{\pm}$, 
which are  independent of 
$X(t)$ and $t$. Solving next
the problem for arbitrary fixed $p_{\pm}$, we determine
 $X(t)$ and the 
 density profile $\rho(\lambda;t)$, 
$\lambda$ being the distance from the edge, $\lambda = X(t) - X$.
 Finally, 
inserting $\rho(\lambda;t)$ into
 Eqs.(1) and (2), we 
find the closure equation, 
which determines $p_{\pm}$ (and hence, $\rm A$)
 self-consistently as functions
of $\rm k_{B} T$ and of the LLI
parameters.
   
The dynamics of a biased RMP in 
a 1D hard-core  gas, placed  in the just described
"shock" configuration, has been studied in Ref.11. 
It was shown that $X(t)$ obeys:
\begin{equation} 
X(t) \; = \; \rm{A} \; \sqrt{\rm{D}_{0} \rm{t}}, 
\end{equation}
where the prefactor $\rm{A}$ 
is defined for $p_{-} > p_{+}$ by
\begin{equation}
\frac{\sqrt{\pi} \; \rm{A}
}{2} \; \exp(\frac{\rm{A}
^{2}}{4}) \;
[ 1 \; + 
\; \Phi(\frac{\rm{A}}{2})] \; = \; 
\frac{\rho \; p_{-}}{p_{-} - p_{+}} \; - \; 1, 
\end{equation}
$\Phi(x)$ being the probability integral.  In the special case
$p_{-} = p_{+}$, $\rm A$ is no longer constant  and 
grows with time as
\begin{equation}
\rm{A} 
\; \propto 
\; \sqrt{2 \; ln(\frac{\rho^2 \omega t}{\pi})},  \qquad{as \; t \to \infty}
\end{equation}
Next, it was found that the  
density profile $\rho(\lambda;t)$ 
past the RMP has a characteristic
 $S$-like shape;  $\rho(\lambda;t)$ 
is almost $constant$ (and different from $\rho$)
in a region of size $\sim X(t)$, 
\begin{equation} 
\rho(\lambda;t) \; = \; (1 \; - \; \frac{p_{+}}{p_{-}}) \; 
[1 \; + \; \frac{A^2 \lambda}{X(t)} \; + O(\frac{A^4 \lambda^2}{X^2(t)}) ],  
\end{equation} 
while for greater $\lambda$, $\lambda \gg X(t)/A^2$, it approaches  $\rho$
exponentially fast. Such a form of $\rho(\lambda;t)$ 
stems from the fact that vacancies propagate
into the gas-phase only diffusively and thus homogenize 
the density distribution 
past the RMP only at scales of order of $X(t)$.
Note, however, that
the total number of particles is conserved, i.e., $lim_{L \to \infty} L^{-1} \int_0^L
d\lambda \; \rho(\lambda;t) = \rho$.

Turning now back to the 2D problem under study,  
we recall  that  $p_{\pm}$  are 
not arbitrary parameters, but their  
 values are determined by the density distribution
past the edge.
Inserting Eq.(7) into the 
Eqs.(1) and (2), we find the self-consistent 
closure equation for 
 $p_{+}/p_{-}$:
\begin{equation}
\frac{p_{+}}{p_{-}} \; = \; exp( - \; \beta \; 
\sigma \; \gamma_{e}); \; \gamma_{e} \; = \; 
(1 - \frac{p_{+}}{p_{-}}) \; \frac{\rm{U}
_{0}(\rm T) \; \delta}{2 \sigma},
\end{equation}
\begin{equation}
\delta \; =  \; \sigma^{\rm n}  
 \sum_{\bf r'', r'' \neq r_{\pm}} \{
\rm \frac{1}{|\bf r_{-} - \bf r''|^{n}} \; - \;
\frac{1}{|\bf r_{+} - \bf r''|^{n}}\},
\end{equation}
where 
$\bf r_{\pm}$ denote the $2D$ vectors $(X(t) \pm \sigma,Y)$. 

Therefore, we find that the mean 
displacement of the 
monolayer edge obeys Eq.(4),
in which $\rm{A}$ is related to $\rm{U}
_{0}(\rm T)$, 
 $\rho$
 and $\beta$
through  Eqs.(5) and (8).  
Below we present some analytical estimates of ${\rm A}(\varepsilon)$,
where $\varepsilon = \beta \; \rm{U}
_{0}(\rm T) \; \delta/2$ is a critical dimensionless parameter.

We find that depending on the value of the parameter $\varepsilon$
four different regimes can be
observed:\\
(1) When $\varepsilon \in [0;1]$ the only solution of   
Eq.(8) is $p_{+}/p_{-} = 1$, 
which means that here
  $\gamma_{e} \equiv 0$
and the monolayer behaves as  
an ideal surface gas.  
In this regime  
 $X(t) \sim \sqrt{t \; ln(t)}$ and the density $\rho(\lambda;\rm t)$ 
changes rapidly with the distance $\lambda$ 
from the edge being equal to $\rho$ for $\lambda \to \infty$ and to zero
for $\lambda = 0$.\\
(2) When $\varepsilon \in ]1;\varepsilon_{c}[$, where 
$\varepsilon_{c} =  -  ln(1 - \rho)/\rho$, the prefactor $\rm A$
is positive and finite.  
Here, $X(t) \sim 
\sqrt{ t}$ and the monolayer 
also wets the solid. 
The edge tension $\gamma_{e} > 0$
 and vanishes as $\gamma_{e} \sim ({\rm T}_{b} - {\rm
T})$ when  $\rm T \to  \rm T_{b}$; $\rm T_{b}$ is thus
the critical  
temperature of the surface 
gas-liquid transition, which is
defined implicitly
by equation $\rm T_{b} = \rm U_{0}(T_{b}) \; \delta/2 k_{B}$, ($\varepsilon = 1$). 
In this regime $\rm{A}$ diverges
when $\varepsilon \to + 1$, 
$\rm{A} 
 \approx 
 \sqrt{ln(\rho/(\varepsilon - 1))}$, 
and vanishes
 when $\varepsilon \to \varepsilon_{c}$, 
$\rm{A} 
 \approx   (1 - \rho) \; (\varepsilon_{c}
 - \varepsilon)/(1 - (1 -
\rho) \varepsilon_{c})$. The density $\rho(\lambda;\rm t)$ changes smoothly
from the unperturbed value $\rho$ to the value at the edge 
$exp(-\beta \sigma \gamma_{e})$, which is close to $\rho$.\\  
(3) At $\varepsilon = \varepsilon_{c}$ the prefactor $\rm{A}$ is exactly 
equal to zero and 
 the monolayer partially wets the substrate.  
Hence, we denote $\varepsilon = \varepsilon_{c}$ 
as the point of the
wetting/dewetting transition for  the monolayer. 
The corresponding critical temperature
is determined by  $\rm{T}
_{w/dw} =   \rm{U}
_{0}(\rm{T}
_{w/dw}) \; \delta/(2 \; k_{B} \; \varepsilon_{c})$ and 
depends on the coverage $\rho$.
  We note that
$\rm{T}
_{w/dw}$ and  $\rm{T}_{b}$
are simply related
to each other. 
When  $\rm{U}
_{0}(\rm{T})$ is independent of $\rm{T}$  (say, for 
 London-type LLI) one has $\rm{T}
_{w/dw} =  \; \rm{T}_{b}/\varepsilon_{c}$.
Actually, the inset in Fig.2 
displays the $\rho$-dependence of the ratio $\rm{T}
_{w/dw}/\rm{T}
_{b}$  ($= 1/\varepsilon_{c}$) 
for this case.
For the Keesom-type interactions, when $U_{0}(\rm{T}) \sim 1/\rm{T}$, $\rm{T}
_{w/dw} = \rm{T}
_{b} /\sqrt{\varepsilon_{c}}$. 
We finally remark that the relation between $\varepsilon$
and
$\varepsilon_{c}$ distinguishes
whether it is favorable, at given physical conditions,  
to have a monolayer with coverage $\rho$
on the solid surface or not. 
Consequently, 
knowing $\varepsilon_{c}$
and the density distribution in
a sessile drop with the respect to the 
height above the substrate,
one can predict the number
of superimposed layers in the
"terraced wetting" regime.\\ 
(4) For $\varepsilon > \varepsilon_{c}$
 the prefactor $\rm{A} 
< 0$ and the monolayer dewets  
from the substrate. Here, a jammed region (where 
$\rho(\lambda;t) > \rho$, Eq.(7)) of size $\sim X(t)$ appears, which impedes
the motion of the retracting edge; 
the $\varepsilon$-dependence of $\rm{A}$ is thus very weak,
 being
strongly limited  by the diffusive squeezing 
out of "voids" at progressively larger and larger
scales.  
In fact, for sufficiently large $\varepsilon$
 one may expect that
  the
 dewetting process
will be accelerated 
by thickening of the monolayer,
 as it is suggested in  Ref.6.

\vspace{0.3cm}

The authors thank S.F.Burlatsky, T.Blake,  E.Rapha\"el and  J.L.Lebowitz
 for  helpful discussions. 
Financial support from FNRS, the PROCOPE-grant,  COST Project  D5/0003/95
and EC Human and Capital Mobility Program CHRX-CT94-0448-3
  is gratefully
acknowledged.

\end{document}